\documentclass[aps,prd,showpacs,amsmath,amssymb,reprint]{revtex4-1} 
\usepackage{graphicx}
\usepackage{color}
\usepackage{ulem}

\newcommand{\sig}{\sigma}
\newcommand{\Ain}{A^{(\text{in})}_m}
\newcommand{\Aout}{A^{(\text{out})}_m}
\newcommand{\tnew}{\tilde{t}}
\newcommand{\phinew}{\tilde{\phi}}
\newcommand{\beq}{\begin{equation}}
\newcommand{\eeq}{\end{equation}}

\begin{document}

\title{Scattering by a draining bathtub vortex}

\author{Sam R. Dolan}
 \email{s.dolan@sheffield.ac.uk}
  \affiliation{%
Consortium for Fundamental Physics, School of Mathematics and Statistics, University of Sheffield, Hicks Building, Hounsfield Road, Sheffield S3 7RH, United Kingdom.
}%

\author{Ednilton S. Oliveira}
 \email{ednilton@ufpa.br}
  \affiliation{
  Faculdade de F\'isica, Universidade Federal do Par\'a, 66075-110, Bel\'em, PA, Brazil.
 }

\date{\today}

\begin{abstract}
We present an analysis of scattering by a fluid-mechanical `black hole analogue', known as the draining bathtub (DBT) vortex: a two-dimensional flow which possesses both a sonic horizon and an ergoregion. We consider the scattering of a plane wave of fixed frequency impinging upon the vortex.
At low frequency, we encounter a modified Aharonov-Bohm effect. At high frequencies, we observe regular `orbiting' oscillations in the scattering length, due to interference between contra-orbiting rays. We present approximate formulae for both effects, and a selection of numerical results obtained by summing partial-wave series. Finally, we examine interference patterns in the vicinity of the vortex, and highlight the prospects for experimental investigation.
\end{abstract}

\maketitle

% Paper Plan:
%
% Introduction:
%   

\section{Introduction}

A standard question in physics is: how do we deduce the properties of something that is too tiny, remote, transient or inaccessible to be observed directly? A common answer is: by observing the way that it scatters.  Depending on the situation, the background irradiation may be natural (e.g.~as in a rainbow), or contrived (e.g.~as in ultrasound imaging). Either way, one is faced with the problem of inferring the properties of scattering bodies from the flux of scattered radiation in the far-field.
%It is no surprise that scattering processes are of interest in subjects as diverse as particle physics, geophysics, astrophysics and condensed matter theory. 

Black holes are an intriguing prediction of General Relativity, which are implicated in some of the most violent astrophysical processes in the universe. For example, black holes accreting matter are powerful emitters of electromagnetic radiation. Yet, most stellar-remnant black holes in our galaxy are likely to be in a quiescent state. It is possible that such black holes may one day be studied via the background light and radiation that they scatter \cite{Tamburini, Dolan, SYG}, for example through observations of microlensing events. 

A rotating black hole, described by the Kerr metric, exhibits two key features: an event horizon (a null hypersurface which acts as a one-way membrane \cite{Membrane}), and a stationary limit surface (which circumscribes an \textit{ergoregion}, or ergosphere). Within the ergoregion, all observers are necessarily co-rotating with the black hole~\cite{Chandra}. %Very loosely speaking, (i) the event horizon marks the point where the inward radial flow of spacetime exceeds the speed of light; and (ii) the stationary limit surface occurs where the rate of ``frame-dragging" around the black hole exceeds the same limit. 
The existence of these features may possibly be confirmed by the imprint they leave upon gravitational radiation. However, since we still await ``first-light'' detections of gravitational waves from (e.g.) black hole mergers, such observations may be some way off.

%Given that black holes are (thankfully) both remote and inaccessible, it is no surprise that the scattering of waves by black holes has received some attention.

A more immediate possibility is that such features may be studied in the laboratory, by constructing \textit{black hole analogues}: systems which mimic some key aspects of black holes \cite{Unruh-1981,Visser-1998-cqg}. 
%To mimic an horizon or ergosphere, one must prepare a medium in which the local flow rate may become faster than the speed of perturbations in the flow. 
Many analogue systems have been proposed in acoustics, optics, and condensed matter theory~\cite{Barcelo-Liberati-Visser,Novello-Visser-Volovik}. Perhaps the simplest analogue model to exhibit both an horizon and an ergoregion has been called the \textit{draining bathtub} (DBT) vortex, or, simply, draining bathtub \cite{Visser-1998-cqg, Unruh-Schutzhold-2002}. This is a model of a two-dimensional flow with a sink at the origin, in a fluid which is assumed to be barotropic and inviscid, and in which flux is locally irrotational. The velocity field $\mathbf{v}$ of the background flow, expressed in polar coordinates $\{t,r,\phi \}$, is
\beq
\mathbf{v} =  \frac{C \hat{\phi} - D \hat{r}}{r},
\eeq
where $C$ and $D$ are constants, with $C$ setting the \textit{circulation} and $D$ the \textit{draining} rate of the vortex (N.B. symbols $B$ and $A$ are used elsewhere in the literature \cite{Visser-1998-cqg}).  
The flow in the DBT is everywhere irrotational (${\mathbf \nabla} \times {\mathbf v} = 0$), except at the sink itself. Then, small perturbations $\delta \mathbf{v}$ in the flow may be expressed as the gradient of a potential function, i.e.~$\delta \mathbf{v}  = - \nabla \psi$. In Ref.~\cite{Unruh-1981} it was shown that the linearized Navier-Stokes equations determining the evolution of $\psi$ lead to the Klein-Gordon equation for a scalar field propagating on an \textit{effective spacetime}:
\beq
\Box \psi \equiv \frac{1}{\sqrt{-g}} \partial_\mu \left( \sqrt{-g} g^{\mu \nu} \partial_\nu \psi \right) = 0.
\label{KG}
\eeq
Here the (inverse) metric $g^{\mu \nu}$ and metric determinant $g$ are found from an effective metric $g_{\mu \nu}$ which is algebraically determined by the background flow rate and fluid properties. For the DBT, the effective line element is simply
\begin{eqnarray}
ds^2 & \equiv & g_{\mu \nu} dx^\mu dx^\nu \nonumber \\
 & = & -c^2 dt^2  + \left(dr + \frac{D}{r} dt \right)^2 + \left(r d\phi - \frac{C}{r} dt \right)^2.
 \label{le1}
\end{eqnarray}
%The system may be constructed in the laboratory; yet the effective metric mimics a curved geometry. 
Here $c$ is the speed of sound in the fluid; following other authors, we will assume that $c$ is constant~\cite{Note1}.  

The DBT model outlined above represents a rather particular idealization of the bathtub vortex as commonly understood in fluid dynamics \cite{Andersen-etal}. In the latter context (i) the vortex has a core of non-negligible radius, so the assumptions of inviscid and irrotational flow naturally break down, (ii) the flow is driven by boundary-layer effects described by Ekman theory, (iii) the flow rate does not come close to the speed of sound, and hence neither horizon nor ergoregion will form, and (iv) the dispersion relation is not linear. These features clearly limit any analogy between wave propagation in a black hole spacetime and in a realistic fluid dynamical system. Nevertheless, as there is widespread interest in exploring phenomenology of rotating black holes, there is also widespread interest in devising an experimental setup which comes close to the idealizations of the DBT model. For example, a possible experiment using surface waves in a fluid tank was outlined in Ref.~\cite{Unruh-Schutzhold-2002}.

The DBT vortex model possesses a (sonic) horizon at $r_h$ and a stationary limit surface at $r_e $, where
\beq
r_h = D / c, \quad \quad r_e = \sqrt{C^2 + D^2} / c .  \label{rh-re}
\eeq
The simpler nondraining vortex ($D = 0$) has been well-studied in both classical~\cite{Lund} (e.g.~fluid dynamics) and quantum~\cite{Leonhardt} (e.g.~superfluidity in Bose-Einstein condensates) contexts. Phonons propagating on a vortex background are subject to the Aharonov-Bohm (AB) effect~\cite{Aharonov,Berry}, focusing with spherical aberration, and frame-dragging effects~\cite{Visser-Weinfurtner-2005}.
In the DBT, due to the presence of a draining component of the flow giving rise to an horizon, additional phenomena will occur such as quasinormal ringing~\cite{Cardoso-Lemos-Yoshida-2004,Leandro}, superradiance~\cite{Berti-Cardoso-Lemos-2004, Basak-Majumdar1, Basak-Majumdar2,Richartz_etal-2012},
and absorption~\cite{odc}. 

 % Frame-dragging becomes most extreme within the ergosphere in which the fluid is moving faster than the speed of sound.

%We show here that scattering by a DBT can be understood and
%analyzed via the acoustic spacetime approach (which is formally
%equivalent to the linearized hydrodynamic equations, with certain
%assumptions). 

In this paper, we will consider the scattering of monochromatic `planar' waves, with incident wavelength $\lambda = 2 \pi c / \omega$, where $\omega$ is the wave's angular frequency (more accurately, we consider the 2D analogue of planar waves: linear waves with straight wavefronts).  Scattering by the DBT is characterized by two just dimensionless quantities,
\beq
\alpha \equiv \frac{\omega C}{c^2},  \quad \quad \beta \equiv \frac{\omega D}{c^2} .  \label{alp-bet-def}
\eeq
Our investigation builds upon two recent works. In Ref.~\cite{odc} the absorption of planar waves by a DBT vortex has been studied. In Ref.~\cite{doc2}, the low-frequency scattering process in a DBT vortex was analyzed, and it was shown that the standard AB effect \cite{Aharonov, Berry}, due to circulating flow, is modified by the presence of a sonic horizon. This has been called the $\alpha\beta$ effect (see also Ref.~\cite{Anacleto-2012}). In this work, we complete that line of inquiry by conducting a comprehensive study of the scattering of planar waves impinging upon a DBT vortex.

%In a separate study, we examined the scattering of waves by a  different analogue: the so-called canonical acoustic hole~\cite{doc}. That system is spherically-symmetric, so does not exhibit an ergoregion. We found that the wave scattering patterns share some characteristics with the Schwarzschild black hole, such as glory oscillations~\cite{Matzner-1985}. The draining vortex, on the other hand, is not symmetric, and may exhibit some of the richer phenomenology associated with the Kerr solution.

The remainder of the paper is organized as follows. In
Sec.~\ref{sec:theory}, we review and develop the relevant theory for the DBT vortex; we examine geodesics (\ref{subsec:geodesics}) and perturbations (\ref{subsec:pert}) on the effective spacetime, and review some concepts of planar-wave scattering in two spatial dimensions (\ref{subsec:sca2D}). In Sec.~\ref{sec:analytic}, we develop our analysis of low-frequency scattering via the Born approximation and the AB effect (\ref{subsec:low-freq}); and high-frequency scattering via semiclassical approximation (\ref{subsec:high-freq}) with application to orbiting oscillations. In Sec.~\ref{sec:results} we describe a numerical method for computing phase shifts and evaluating partial-wave series. We present a selection of numerical results, and we compare them with the approximations of Sec.~\ref{sec:analytic}. In Sec.~\ref{subsec:near-field} we investigate the properties of the interference pattern in the near-field. We conclude with a discussion in Sec.~\ref{sec:conc}. In the following sections, we set the speed of 
sound to unity, $c = 1$.

\section{\label{sec:theory}Foundations}
% In this section, we recap and develop some essential theory for the DBT vortex, and for monochromatic planar-wave scattering in 2D.

\subsection{Effective spacetime}

The DBT spacetime, described by Eq.~(\ref{le1}), may be alternately expressed if we first make a change of variables to a new coordinate system \cite{Basak-Majumdar2, Berti-Cardoso-Lemos-2004},
\begin{eqnarray}
d \tnew &=& d t - \frac{Dr}{r^2-D^2} dr, \\
d \phinew &=& d \phi - \frac{CD}{r(r^2-D^2)} dr,
\end{eqnarray}
with $\phinew(r \rightarrow \infty) = \phi$.

In the new coordinate system, the line element takes the form
\beq
ds^2 = - g(r) d \tnew^2 + f(r)^{-1} dr^2 - 2C d \phinew d \tnew + r^2 d \phinew^2,  \label{metricnew}
\eeq
where
\begin{equation}
g(r) = 1 - \frac{C^2+D^2}{r^2}, \qquad \text{and} \qquad
f(r) = 1 - \frac{D^2}{r^2} .
\end{equation}
Note that the laboratory coordinates ($t,r,\phi$) are the analogue of the `ingoing Kerr' coordinates, whereas the new coordinates ($\tnew, r, \phinew$) are the analogue of `Boyer-Lindquist' coordinates (used to describe a rotating black hole spacetime). 

In the following sections, we work with the alternative coordinate
system (\ref{metricnew}) but, for clarity, we drop the tilde ($\tilde{}$) notation. 

\subsection{Geodesics\label{subsec:geodesics}}
In an idealized system without dispersion, very high-frequency perturbations will propagate along the null geodesics of the effective spacetime. Let us begin, therefore, by considering null geodesics in the context of scattering, expanding on the treatment of Sec.~IVB in Ref.~\cite{odc}. The geodesic analysis reveals characteristic properties of the effective acoustic spacetime, such as the critical orbits, that play important roles in the scattering process even at moderate frequencies (Sec.~\ref{subsec:high-freq}). We note that short-wavelength fluid perturbations will most likely obey a non-linear dispersion relation~\cite{Unruh-1995}, and this has to be taken into account, for instance, in the study of acoustic Hawking radiation~\cite{Weinfurtner_etal-2011,Rousseaux_etal-2010,Jannes_etal-2011}. 

The metric associated to Eq.~(\ref{metricnew}) is independent of $t$ and $\phi$.
Hence, there are two Killing vectors giving rise to two constants of geodesic motion, $E$ and $L$, corresponding
to energy and angular momentum, respectively:
\begin{eqnarray}
E  & \equiv & -u_{t}  = g(r) \dot{t} + C \dot{\phi} ,\nonumber\\
L & \equiv & \phantom{-} u_{\phi} = r^2   \dot{\phi} - C \dot{t} ,
\label{constants-of-motion}
\end{eqnarray}
where $u^\mu = dx^\mu / d\nu$ is the null geodesic four-velocity, and the overdot denotes differentiation with respect to an affine parameter $\nu$. It is straightforward to rearrange
Eq.~(\ref{constants-of-motion}) to find
\beq
\dot{t} = \frac{E - CL/r^2}{f} , \qquad \dot{\phi} =
\frac{L}{r^2} + \frac{C E -LC^2/r^2}{fr^2}.  \label{tdot-phidot}
\eeq
Now, substituting (\ref{tdot-phidot}) into the defining equation for null rays ($g_{\mu \nu} \dot{x}^\mu \dot{x}^\nu= 0$) we obtain the `energy' equation
\beq
\dot{r}^2 = \left( E - \frac{CL}{r^2} \right)^2 - \left(1 - \frac{D^2}{r^2} \right) \frac{L^2}{r^2}.
\label{energy-eq}
\eeq
The orbital equation may be written in the following fashion
\begin{eqnarray}
\left(\frac{d u}{d \phi} \right)^2 & = & \frac{f^2}{(C + gl)^2} \left[ 1 - l(l+2C) u^2 \right. \nonumber\\ && \left. \quad + l^2(C^2 + D^2) u^4 \right],  \label{orb-eq}
\end{eqnarray}
where $u \equiv 1/r$ and $l$ is the \textit{specific angular momentum},
\beq
l \equiv L/E .   \label{spec-ang-mom}
\eeq
The impact parameter $b$ for a geodesic incident from spatial infinity is related to $l$ via $b = l + C$.
When the background is rotating ($C \neq 0$), $L$, $l$ (and thus
$b$) may take either sign: $l$ is positive for a co-rotating ray and negative
for a counter-rotating ray.

\subsubsection{Critical Orbits}

Null circular orbits occur at radii
where simultaneous conditions $\dot{r} = 0$ and $\ddot{r} = 0$ are
satisfied. There are two critical values of angular momentum, $l_c^{+}$ and $l_c^{-}$, corresponding to co-rotating and counter-rotating null rays incident from infinity which end on
(unstable) circular orbits at $r_c^{+}$ and $r_c^{-}$, respectively,
given by
\begin{eqnarray}
l_c^{\pm} &=& \pm 2 \sqrt{D^2 + C^2} - 2 C ,  \nonumber \\
r_c^{\pm} &=& \left( \sqrt{D^2 + C^2} | l_c^\pm |
\right)^{1/2} ,  \label{crit-orb-params}
\end{eqnarray}
and
$b_c^{\pm} = l_c^{\pm} + C $.
Note that $|b^{+}_c| \le |b^{-}_c|$ and $|r^{+}_c| \le |r_c^{-}|$ (for $C \ge 0$). In other words, the co-rotating rays may approach closer to the vortex than the counter-rotating rays without being absorbed, as
can be seen in FIG.~\ref{fig-trajectories}.

\subsubsection{Scattering Angle}
Let us now consider those null geodesics incident from spatial infinity that are scattered to infinity rather than being absorbed ($b < b_c^{-}$ or $b > b_c^{+}$). By solving Eq.~(\ref{orb-eq}), we find that the scattering angle $\Theta$ may be expressed in terms of complete elliptic integrals of the first ($K$) and third ($\Pi$) kinds~\cite{Gradshteyn}, as
\begin{eqnarray}
\Theta + \pi & = & \frac{2\sqrt{C^2 + D^2}}{D^2 u_1}
\left[ K\left(\frac{u_0}{u_1}\right) \right. \nonumber\\
&& \left . + \frac{C(D^2 - l C)}
{l (C^2 + D^2)} \Pi\left(D^2 u_0^2, \frac{u_0}{u_1}\right) 
\right],   \label{scattering-angle}
\end{eqnarray}
%\red{[please check]} checked once
%
where $u_0$ and $u_1$ are the positive roots of the polynomial
\beq
u^4 - \frac{1 + 2C/l}{D^2+C^2} u^2 + \frac{1}{l^2(D^2+C^2)} = 0 ,  \label{upoly}
\eeq
such that $0 < u_0 \le u_1$.

\begin{figure}[htb]
 \includegraphics[width=8cm]{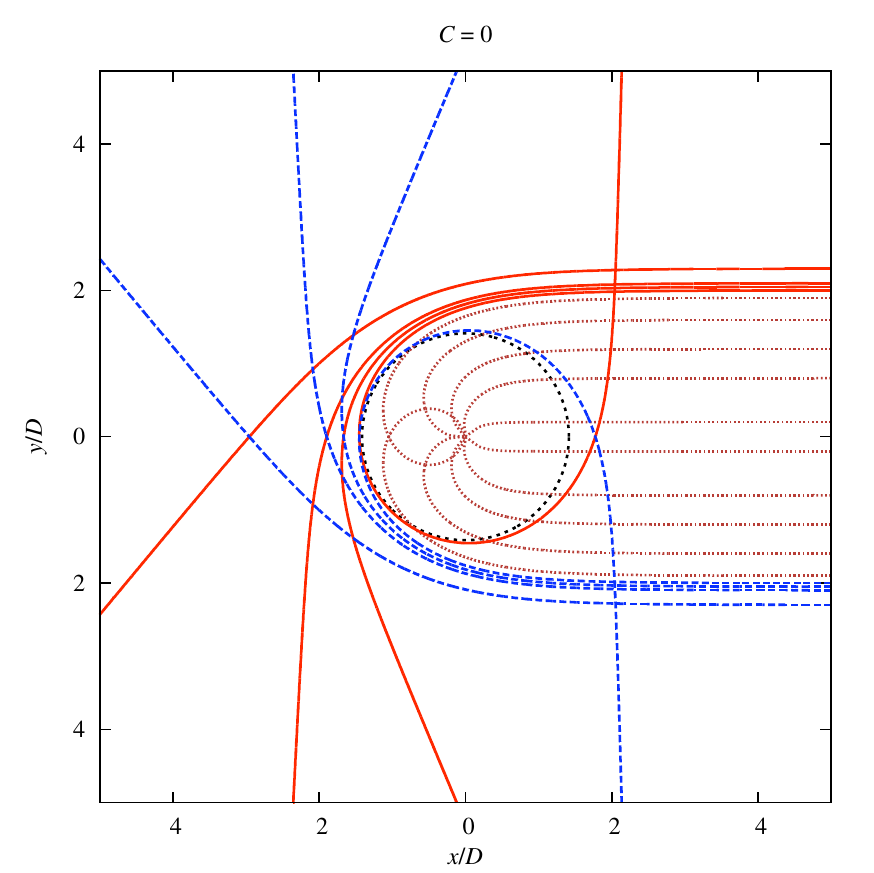}
 \includegraphics[width=8cm]{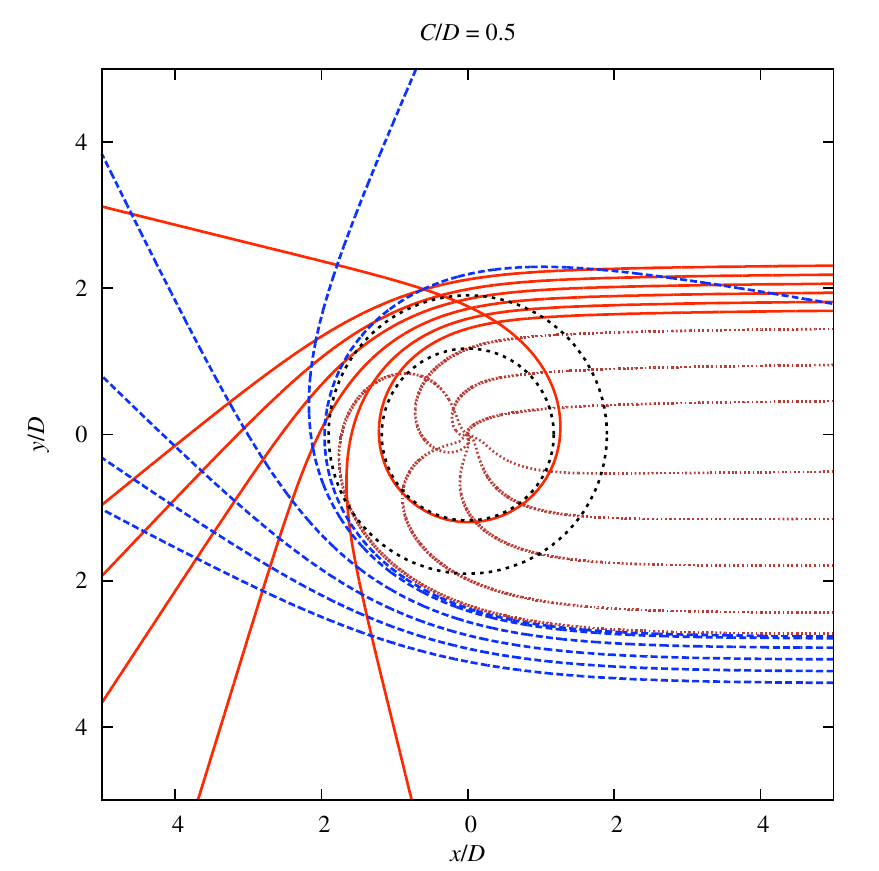}
 \caption{Null geodesics impinging on a draining bathtub for the static case (top) and rotating case with $C/D = 0.5$ (bottom). Co-rotating, scattered geodesics are represented in solid (red) lines, whereas counter-rotating scattered geodesics are represented by dashed (blue) lines. Dotted (brown) lines represent absorbed geodesics. Dotted (black) circular lines represent the null critical orbits. Note that the co-rotating
 geodesics may pass closer to DBT vortex than counter-rotating geodesics, without being absorbed.
 }
 \label{fig-trajectories}
\end{figure}

For rays with large impact parameters, $|b| \gg |C|,|D|$, the scattering angle can be approximated by
% \red{[SD: I have written this equation in terms of $b$ rather than $l$, as in the AB paper]}
\beq
\Theta \approx \frac{3\pi}{4 b^2}\left(C^2 + D^2 \right) - \frac{\pi C}{b^3} \left(C^2 + D^2\right) + \mathcal{O}(b^{-4}) .
\label{weakfield}
\eeq
%\beq
%\Theta \approx \frac{3\pi}{4 l^2}\left(C^2 + D^2 \right) - \frac{5 \pi C}{2 l^3} \left(C^2 + D^2\right) + \mathcal{O}(l^{-4}) .
%\label{weakfield}
%\eeq
Note that the dominant term is the same for co-rotating and counter-rotating geodesics, but the sub-dominant term depends on the sign of $C/b$.

Rays that pass close to the unstable null orbits (with impact parameters $b \gtrsim b_c^{+}$ and $b \lesssim b_c^{-}$) may be scattered through large angles. Close to the critical impact parameter, we find that the scattering angle depends logarithmically on $b-b_c$. For co-rotating rays, 
\begin{eqnarray}
\Theta + \pi & \approx & - \frac{r_c^+}{2D} \left( \sqrt{1+ \frac{C^2}
{D^2}} + \frac{C}{D} \right) \ln\left[ \frac{(b-b_c^{+}) {r_c^+}^2}{64 l_c^+ (D^2+C^2)}
\right] \nonumber\\
&& - \frac{C}{D} \ln\left( \frac{r_c^+ + D}{r_c^+ - D} \right).
\end{eqnarray}
%\red{Need to check.} checked once
For a similar expression in the black hole case, see Ref.~\cite{Darwin}.

Geodesics on static and rotating DBT vortices are illustrated in FIG.~\ref{fig-trajectories}.

\subsection{Perturbations of the DBT flow\label{subsec:pert}}

Making the separation ansatz 
\begin{equation}
\psi = [R_{\omega m}(r)/
\sqrt{r}] \exp[ i (m \phi - \omega t)]
\end{equation}
in the Klein-Gordon equation (\ref{KG}) leads to the radial equation
\begin{eqnarray}
\frac{d^2 R_{\omega m}}{d r_*^2} && + \left\{ \left(\omega - \frac{Cm}
{r^2} \right)^2 - \right.  \nonumber\\
&& \left. \frac{f}{r^2} \left[\left(m^2 - \frac{1}{4} \right)
+ \frac{5D^2}{4r^2} \right] \right\} R_{\omega m} = 0,
\label{radialeq}
\end{eqnarray}
where the tortoise coordinate $r_*$ is defined by
\beq
\frac{d r_*}{dr} = f^{-1} \quad \Longrightarrow \quad r_* = r + \frac{D}{2} \ln \left| \frac{r-D}{r+D} \right|.
\label{tortoise}
\eeq
It is possible to find the asymptotic behavior of the radial function
$R_{\omega m}(r_*)$ from Eq.~(\ref{radialeq}). In the regime $r \gg r_h$, for a slowly rotating acoustic black hole ($C \ll r$), Eq.~(\ref{radialeq}) reduces to:
\begin{equation}
\frac{d^2 R_{\omega m}}{dr_{*}^{2}} + \left( \omega^2 -
\frac{\nu^2 - 1/4}{r_{*}^2} \right) R_{\omega m} = 0,
\label{asy_eq}
\end{equation}
where $\nu = \sqrt{ \left|m(m+2\alpha)\right|} $. The solution of Eq.~(\ref{asy_eq}) may be written as 
\begin{eqnarray}
R_{\omega m} \approx \sqrt{\frac{\pi\omega r_*}{2}}
\left[A_{m}^{(\text{in})}
e^{-i(\nu +1/2)\pi/2} H_{\nu}^{(1)*}(\omega r_*) \right.
\nonumber\\
\left. + A_{m}^{(\text{out})} e^{i(\nu + 1/2)\pi/2}  H_{\nu}^{(1)} (\omega r_*)
\right],
\label{asy_sol}
\end{eqnarray}
where $A_{m}^{(\text{in/out})}$ are coefficients depending on $\omega$ and $m$, and  $H_{\nu}^{(1)}$ is a Hankel function~\cite{Gradshteyn}.
Using the asymptotic form of the Hankel function, it follows that, in the far-field,
\begin{equation}
R_{\omega m} \sim A_{m}^{(\text{in})} e^{-i\omega r_*} +
A_{m}^{(\text{out})} e^{i\omega r_*}.
\label{inf_sol}
\end{equation}
%where we see that $\left|A_{m}^{(\text{out})}/
%A_{m}^{(\text{in})} \right|^2$ is the
%reflection coefficient related to the part of the incident wave which
%goes back to infinity.

Physical solutions of the radial equation (\ref{radialeq})
are subject to an `ingoing' boundary condition at the horizon,
\beq
R_{\omega m} \sim e^{-i \tilde{\omega} r_* } , \quad \text{where}
\quad \tilde{\omega} \equiv  \omega - \frac{mC}{D^2} .
\label{h_bc}
\eeq

%This is enough to analyze the scattering properties, such as the phase shifts and scattering cross section, once these properties depend only on the asymptotic wave behavior. 
%Note that, on this limit, both coordinates systems behave likely.

\subsection{Time-independent scattering theory in two
dimensional space\label{subsec:sca2D}}

Scattering theory in two spatial dimensions is outlined in, e.g.,
Refs.~\cite{Lapidus, Adhikari, Adhikari-Hussein}. We can construct a monochromatic solution $\psi = e^{-i\omega t} \psi(r,\phi)$ by decomposing it into partial waves,
\beq
\psi(r, \phi)  =  \frac{1}{\sqrt{r}} \sum_{m = -\infty}^{\infty} e^{i m \phi} R_{\omega m}(r) .
\eeq
The solution we seek is the superposition of a planar wave propagating in the $+x$ direction, and a scattered component with amplitude $f_\omega(\phi)$, i.e.
\beq
\psi(r, \phi)  =  e^{i \omega x} + f_\omega (\phi) \frac{e^{i\omega r_\ast}}{\sqrt{r}}.
\eeq
Note that (in our convention) $f_\omega$ has dimensions of ${(\text{Length})}^{1/2}$.

Asymptotically, the solutions take the form (\ref{inf_sol}). 
%\begin{equation}
%R_{m \omega} (r) \sim \left(  \Ain e^{-i \omega r_\ast} + \Aout e^{+i \omega r_\ast}  \right)  \frac{e^{i (\omega r_\text{h} / c) \ln r}}{\sqrt{r}}.
%\end{equation}
% We will ignore the logarithmic phase factor , since it affects all modes in the same way. 
The plane wave may be decomposed into
\begin{equation}
e^{i \omega x} = \sum_{m=-\infty}^{\infty} i^m e^{i m \phi} J_m (\omega r),
\end{equation}
where $J_m(\cdot)$ are the Bessel functions of first kind~\cite{Gradshteyn}.
Using this, together with the asymptotic form of the Bessel functions,
one finds that the scattering amplitude is
\begin{equation}
f_\omega (\phi) = \left( \frac{1}{2 i \pi \omega} \right)^{1/2} \sum_{m=-\infty}^\infty \left( e^{2i \delta_m} - 1 \right) e^{i m \phi},   
\label{amp}
\end{equation}
where the phase shifts are defined by
\begin{equation}
e^{2 i \delta_m} = i (-1)^{m} \Aout / \Ain ,  \label{phaseshift-def}
\end{equation}
and the coefficients $\Aout$ and $\Ain$ are found from the asymptotic form (\ref{asy_sol}) [or (\ref{inf_sol})].

Following Ref.~\cite{com}, by constructing a conserved current from the
Klein-Gordon equation, the
absorption cross section (with dimension of length) is
\begin{equation}
\sig_{\text{abs}} = \frac{1}{\omega} \sum_{m = -\infty}^\infty \left(1
- \left| e^{2i \delta_m} \right|^2 \right).
\end{equation}
The absorption of planar waves by a DBT was investigated in detail % by the present authors
in Ref.~\cite{odc}.

The differential scattering length $d\sigma / d \phi$ (which we will also call the `scattering cross section') follows directly from the scattering amplitude, 
\begin{equation}
\frac{d \sig}{d \phi} = \left| f_\omega (\phi) \right|^2.  \label{dsig-def}
\end{equation}

\section{Analytical Development\label{sec:analytic}}

\subsection{Low-frequency scattering: The Aharonov-Bohm effect\label{subsec:low-freq}}

In Ref.~\cite{doc2} it was shown that, in the limit $|m| \gg \sqrt{\alpha^2 + \beta^2}$ (with
$\alpha$ and $\beta$ being the dimensionless couplings defined in Eq.~(\ref{alp-bet-def})), it is possible to find the approximate form of the phase shifts (defined in Eq.~(\ref{phaseshift-def})) via the Born approximation, namely
\beq
\delta_m = - \frac{\pi \alpha}{2} \frac{m}{|m|} +
\frac{3 \pi (\alpha^2 + \beta^2)}{8 |m|} -
\frac{5 \pi \alpha (\alpha^2 + \beta^2) }{8 m^2} \frac{m}{|m|}.
\label{phaseshift-wkb}
\eeq
Note that, at leading order, the phase shift depends only on the sign of $m$,
but not on its magnitude. In other words, if the system is
rotating ($\alpha \neq 0$), there is a nonzero phase shift even in the limits $m
\rightarrow \pm \infty$. 

We may recover the weak-field deflection angle
(\ref{weakfield}) from the phase shift via the \textit{semi-classical} (SC)
relation~\cite{MTW}
\beq
\Theta = - \frac{d}{dm} \left(2 \delta_m \right),  \label{semi-classical-defl}
\eeq
by making the SC association $l \leftrightarrow m/\omega$. Hence, the second and third terms in (\ref{phaseshift-wkb}) are related, respectively, to the circularly-symmetric deflection in the weak-field, and the lowest-order rotation-dependent correction. The first term in (\ref{phaseshift-wkb}) cannot be interpreted so straightforwardly, since it is not related to a deflection of geodesics. Instead, it is linked to the relative time difference $|\Delta t |= 2\pi C$ accrued by geodesics passing on
opposite sides of the vortex in the weak-field. This leads to an interference in long-wavelength perturbations that is the analogue of the AB effect \cite{doc2}. Note that this term is proportional to $\alpha = \omega C$, whereas the leading deflection term is proportional to $\alpha^2 + \beta^2$. Hence, this term is dominant at sufficiently low frequencies. Fischer and Visser noted an analogous behavior for vortices in Ref.~\cite{Fischer-Visser-2002}, observing that the AB effect is dominant for long wavelength modes $k \ll k_c \equiv 2 \pi \sqrt{\mathcal{R}}$, where $\mathcal{R}$ is the scalar curvature (Ricci scalar) of the acoustic metric,
$
\mathcal{R} = 2 (C^2 + D^2) / r^4 .
$

It is possible to find the low-frequency scattering length directly from the phase shift approximation~(\ref{phaseshift-wkb}). In the limit
$\alpha \gg \alpha^2+\beta^2$, we can write
\beq
\delta_{m} \approx -\frac{\alpha \pi}{2}\frac{m}{|m|} + \mathcal{O}(\omega^2) , \qquad (m \neq 0).
\eeq
In Ref.~\cite{doc2} it was shown that the $m=0$ mode has a simple analytic solution, and that
\beq
\delta_{m=0} = \frac{1}{2} i \pi \beta .
\eeq
The low-frequency scattering amplitude can thus be written as
\begin{eqnarray}
f_\omega (\phi) %&=& \left( \frac{1}{2 i \pi \omega} \right)^{1/2} \sum_{m = -\infty}^{\infty} \left(e^{2 i \delta_m} - 1 \right) e^{i m \phi} \nonumber\\
 &\approx & \left( \frac{1}{2 i \pi \omega} \right)^{1/2} \left\{ \sum_{m=1}^\infty \left[ (e^{-i\pi\alpha} - 1) e^{im\phi}  + \text{c.c.}  \right] + \right. \nonumber \\
 & & \quad \quad \quad \quad \quad  \left. + \, e^{-\pi \beta} - 1  \right\},
\end{eqnarray}
where $\text{c.c.}$ denotes the complex conjugate. 
To lowest order in $\omega$, this generates the scattering cross
section
\begin{equation}
\frac{d \sigma}{d \phi} = \frac{\pi}{2\omega} \frac{[\alpha
\cos(\phi/2) - \beta\sin(\phi/2)]^2}{\sin^2(\phi/2)} + \mathcal{O}
(\omega^2).
\label{ana_sca}
\end{equation}
This implies that the (low-frequency) scattering length is zero at angle $\phi_0 = 2 \arctan (\alpha / \beta)$. 

In the nondraining limit ($\beta = 0$), Eq.~({\ref{ana_sca}) reduces to
\beq
\frac{d\sigma_\text{vortex}}{d\phi} =
\frac{\pi \alpha^2}{2\omega} \cot^2(\phi/2),
\label{low_nondraining}
\eeq
which is found in studies of scattering by a nondraining vortex [\textit{cf}. Eq.~(71) in Ref.~\cite{Fetter}].  Eq.~(\ref{low_nondraining}) may also be compared against the scattering
length for the AB effect [\textit{cf}. Eq.~(23) of Ref~\cite{Berry}]:
\beq
|f|^2 = \frac{1}{2 \pi \omega} \frac{ \sin^2(\pi \tilde{\alpha}) }{ \sin^2(\phi/2) },
\label{AB_sca}
\eeq
where $\tilde{\alpha} = e \Phi / (h c)$ in the quantum-mechanical scenario (with $\Phi$ denoting the magnetic flux, $e$ the electron's charge and $h$ Planck's constant \cite{Aharonov}), or $\tilde{\alpha} = \omega \Omega / (2 \pi c_s^2)$ in the water-wave analogue (with $\Omega$ denoting the vorticity and $c_s^2$ is the product of group and phase velocities \cite{Berry}). 

\begin{figure}[htb!]
\centering
\includegraphics[scale=1]{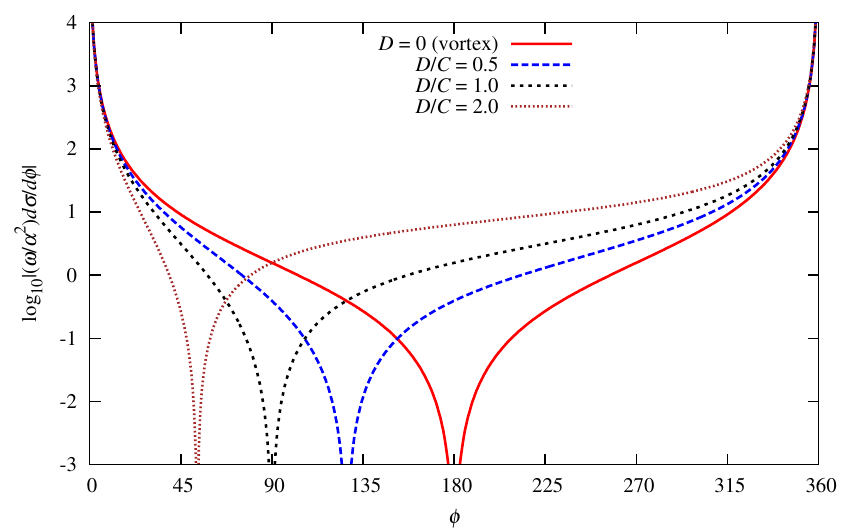}
\caption{Low-frequency vortex scattering for $D = 0$ (nondraining vortex), and for $D/C = 0.5, 1.0, 2.0$. The draining component of flow breaks the symmetry of the scattering length; however, the scattering
behavior near the forward direction does not change. The zero occurs at $\phi_0 = 2 \arctan(C / D)$.}
\label{low_freq-sca}
\end{figure}

Figure~\ref{low_freq-sca} shows the low-frequency scattering length (\ref{ana_sca}) as a function of scattering angle, for different choices of the ratio of circulation $C$ and
draining $D$ parameters. In the non-draining (and also the non-circulating) case, the scattering length is symmetric. The symmetry is broken when draining is switched on, and becomes
stronger as we increase the value of $D/C ( = \beta / \alpha)$. The break in symmetry is
due to the absorption in the isotropic ($m=0$) mode, which is independent of $C$. In the low-frequency limit, the absorption length equals the acoustic hole's circumference, $\sigma_\text{abs} 
\approx 2\pi D$~\cite{odc}. 
%Therefore, in this limit, the absorption length does not depend on the circulating parameter $C$ (although it does at larger frequency values). This is a consequence of the fact that, in this limit, the absorption is only important for the $m = 0$ mode, and the behavior of this mode does not depend on the draining bathtub rotation, as it can be seen by letting $ m = 0 $ in Eq.~(\ref{radialeq}).

\subsection{Higher-frequency scattering: Orbiting\label{subsec:high-freq}}
In 2D, the `classical' scattering length is defined as the density of geodesics passing into unit angle. In the trivial case where each angle $\phi$ is associated with a unique scattering geodesic, the classical scattering length is simply
\beq
\left. \frac{d \sigma}{d\phi} \right|_{cl} =  \left| \frac{d \Theta}{d l}  \right|^{-1} ,   \label{classical-csec}
\eeq
where $\Theta$ is the deflection angle given in Eq.~(\ref{scattering-angle}). 
In the case of the DBT, the association is not unique, and thus interference effects can arise. In particular, there arise regular \textit{orbiting oscillations} \cite{Newton, Adhikari-Hussein, Anninos}, of approximately constant angular frequency $2 \pi / \lambda_\phi$, where the angular width $\lambda_{\phi}$ is derived below in Eq.~(\ref{orbiting-wavelength}). In essence, these oscillations arise from interference between a pair of rays which pass in opposite senses around the DBT vortex. 

\subsubsection{The semi-classical approximation\label{sec:SCorb}}
We may employ a SC approximation, valid in the high-frequency limit, to understand this phenomenon. The key steps in the approximation are (i) making the association $m \sim l \omega$, where $l$ is the specific angular momentum defined in (\ref{spec-ang-mom}), (ii) using the relation between the geodesic deflection function $\Theta$ and the derivative of the phase shift, Eq.~(\ref{semi-classical-defl}), in a stationary phase approximation.

Let us start by writing the expression for the scattering amplitude, after neglecting the on-axis contribution, as
\beq
f_{\omega}(\phi) = \kappa \sum_{m=-\infty}^{\infty} S_m e^{i m \phi} ,
\eeq
where 
\beq
\kappa \equiv (2 i \pi \omega)^{-1/2} \quad \text{and} \quad S_m = \exp( 2 i \delta_m ).  \label{Sm-def}
\eeq
Now we may use the Poisson sum formula \cite{Nussenzveig} to convert the sum into an integral,
\beq
f_{\omega}(\phi) =  \kappa \sum_{n=-\infty}^{\infty} e^{-i \pi n} \int_{-\infty}^{\infty} S_m e^{i m \phi} e^{2i \pi n m} dm .
\eeq
To restrict the range of the integral to positive values of $m$, we may exploit the following symmetry,
\beq
S(-m, \omega, C) = e^{-2 i \pi m} S(m, \omega, -C) ,
\eeq
from which it follows that $\delta(m,\omega,C) = \delta(-m, \omega, C) - \pi m$. 
Thus,
\beq
f_{\omega}(\phi) =  \kappa \sum_{n=-\infty}^{\infty} e^{-i \pi n} \int_{0}^{\infty} \left[ e^{i \xi^+_n(m)} + e^{i \xi^-_n(m)} \right] dm , \label{f-Poisson}
\eeq
where the prograde ($+$) and retrograde ($-$) phases are
\beq
\xi^{\pm}_n(m) = 2 \delta(m, \omega, \pm C) \pm m \phi + 2 \pi n m .
\eeq
To evaluate such integrals in the high-frequency regime (where $\xi^\pm_n$ are rapidly-varying functions of $m$), we may use the Stationary Phase Approximation (SPA),
\beq
\int_0^\infty e^{i \xi (m)} dm \approx  \left( \frac{2 i \pi}{\xi^{\prime \prime}(\bar{m})} \right)^{1/2} e^{i \xi(\bar{m})},
\label{SPA}
\eeq
where $\bar{m}$ is the point of stationary phase, satisfying
\beq
\xi^\prime(\bar{m})  =   0  \label{stationary-phase}
\eeq
(here ${}^\prime$ denotes differentiation with respect to $m$). Let us now recall the SC relationship between the phase shift and the deflection function, given in Eq.~(\ref{semi-classical-defl}). 
The stationary phase condition (\ref{stationary-phase}) is equivalent to
\beq
\Theta^{+} = 2 \pi n + \phi , \quad \text{and} \quad \Theta^{-} = 2 \pi n - \phi
\label{Theta_pm}
\eeq
where $\Theta^{\pm} = \Theta(l = m / \omega, \pm C)$. That is, we may associate each term in Eq.~(\ref{f-Poisson}) with either a prograde ($+$) or retrograde ray ($-$) passing $n$ times around the vortex. 

To lowest order, the orbiting oscillations are due to the interference between two principal rays: the first prograde ray passing through angle $\phi$ (where $0 < \phi < 2 \pi$) and the first retrograde ray that passes through angle $2\pi - \phi$. That is,
\beq
f_{\omega C}(\phi) \approx \kappa \int_0^{\infty} \left[ e^{i \xi^{+}_0(m)} - e^{i \xi_{1}^{-}(m)} \right] dm .
\eeq

Summing the stationary-phase contributions from the two rays leads to the result
\beq
\frac{d \sigma}{d \phi} \approx \frac{d \sigma}{d \phi}^{+} + \frac{d \sigma}{d \phi}^{-}  +  I .  \label{sig-semiclassical}
\eeq
Here the former terms denote the `classical' contributions from the rays, 
\beq
\frac{d \sigma}{d\phi}^\pm =  \left| \frac{d \Theta^{\pm}}{d l}  \right|^{-1},
\label{sig-pm}
\eeq
where we have used $\xi^{\pm \prime \prime}_{n} = -\omega^{-1} d \Theta^{\pm} / dl$ (which follows from the SC relation $m \sim l \omega$). The latter term is due to interference between the two dominant contributions to the scattering amplitude, and is given by
\beq
I = -2 \left| \frac{d \Theta^{+}}{dl} \cdot \frac{d \Theta^{-}}{dl} \right|^{-1/2} \cos( \xi^{+}_0 - \xi^{-}_{1} ) .  \label{eqI}
\eeq
To interpret this equation, let us consider the dependence of $\xi^{+}_0$ on the angle $\phi$:
\beq
\xi^{+}_0(\phi)  =  2 \delta( \bar{m}(\phi), \omega, X) + \bar{m}(\phi) \phi.
\eeq 
It follows immediately that $d \xi^{+}_0 / d\phi = \bar{m}^{+}$ (due to the stationary phase condition) and hence 
\beq
-\cos( \xi^{+}_0 - \xi^{-}_1 ) \approx  \cos [ 4 \omega r_e ( \phi - \pi - \chi ) ] ,   \label{Dphi}
\eeq
so that the angular wavelength of the orbiting oscillations is simply
\beq
\lambda_{\phi} =  \frac{\pi}{2 \omega r_e} ,   \label{orbiting-wavelength}
\eeq
where $r_e$ is the radius of the ergoregion, Eq.~(\ref{rh-re}). 
The remaining challenge is to deduce the $C$-dependent angular offset, $\chi$, which is zero in the non-rotating case, $\chi(C=0) = 0$. We expect positive interference where the path difference between co- and counter-rotating rays is an integer multiple of the wavelength. To find $\chi$, we compute the time difference $\Delta t$ between contra-rotating rays scattered into the backward direction. In Appendix \ref{appendix:offset} we give details of this calculation, obtain numerical values for $\chi$, and derive a simple approximation, $\chi \approx 18.08 C / (4 r_e)$ which turns out to be remarkably powerful. In Sec.~\ref{sec:results} we compare the SC prediction against numerical results (see e.g.~FIG.~\ref{fig-semiclassical}). 

\subsubsection{The Complex Angular Momentum method\label{subsec:cam}}
An alternative way to understand the orbiting phenomena is provided by the Complex Angular Momentum (CAM) method, which makes a link between the orbiting oscillations and the poles of the scattering matrix $S_m$ in the complex-$m$ plane, i.e.~the \textit{Regge poles} (RPs). Regge poles occur at \textit{complex} angular momenta $m_{\omega n}$ where $\Ain(m_{\omega n}, \omega) = 0$. Some time ago the CAM method was successfully applied to investigate scattering by a Schwarzschild black hole \cite{Andersson-1994}. More recently the CAM method was extended to treat absorption \cite{DEFF, DFR, DFR2}. In Ref.~\cite{Leandro} the CAM method was used to calculate a high-frequency approximation for the orbiting oscillations. The key results are given in Eq.~(77)--(81) and FIG.~9 of Ref.~\cite{Leandro}. 
% In Sec.~\ref{subsec:scattering-length}, we compare the CAM approximation for orbiting with numerical results from the partial-wave method, and find a good agreement.

\section{\label{sec:results}Numerical Results}
To obtain results for scattering lengths for intermediate values of the frequency of the perturbation,
we used a numerical method. In the following subsections, we describe the method, present a selection of numerical results, and validate them against the approximations of Sec.~\ref{sec:analytic}.

\subsection{Method}

We start with a series expansion at the horizon of the form,
\beq
R_{\omega m} \sim e^{-i \tilde{\omega} r_* }
\sum\limits_{k=0}^{\infty}
a_k(r - r_h)^k,
\label{h_exp}
\eeq
where $a_k$ are coefficients which are straightforward to determine analytically. We use this as the initial condition, and integrate the radial equation (using a 4th order Runge-Kutta scheme) into the large-$r$ regime. Here, we match the numerical solution onto a suitable asymptotic form to determine $A_m^{(\text{in/out})}$ and hence the phase shift. This can be done in two different ways: (i) by matching onto the Hankel functions through Eq.~(\ref{asy_sol}), and
(ii) by matching onto an asymptotic series of the form
\begin{equation}
R_{\omega m}(r) \sim A_{m}^{(\text{out})} e^{i\omega r_*}
\sum\limits_{k=0}^{\infty}\frac{b_k}{r^k} + A_m^{(\text{in})}
e^{-i\omega r_*}\sum_{k=0}^{\infty} \frac{\bar{b}_k}{r^k}, 
\label{series_exp}
\end{equation}
where the $b_k$ coefficients can be found
analytically. We have compared the results from using both methods, and obtained
excellent agreement. 

\subsubsection{Series convergence}
The partial-wave series for the scattering amplitude, Eq.~(\ref{amp}), does not converge. 
%For $C \neq 0$, the AB effect implies that terms in the series do not tend to zero in the large-$m$ limit (a consequence of Eq.~(\ref{phaseshift-wkb})).  
This is unsurprising, since the amplitude diverges in the forward direction. A similar problem occurs in Coulomb scattering \cite{YRW}, or in scattering by a Schwarzschild black hole \cite{Andersson-Thylwe} (but not in scattering by the canonical acoustic hole \cite{doc}). The convergence rate of the series can be improved by adapting a trick used by Yennie \textit{et al}. in Ref.~\cite{YRW} (see also Refs.~\cite{Dolan, Dolan-Doran-Lasenby-2006}). We use the following argument: If some quantity $X$ has a series representation,
\beq
X \equiv \sum_{m = -\infty}^{m = \infty} X_m e^{i m \phi},
\eeq
then
\begin{eqnarray}
\left( 1 - \cos \phi \right) X &=&
 \sum_{m = -\infty}^{\infty} \left[ X_{m} - \tfrac{1}{2} \left( X_{m+1} + X_{m-1} \right) \right] e^{i m \phi},
\nonumber
\end{eqnarray}
and the latter series is more convergent (in our case) than the former. 
By iterating $n$ times, we may compute the scattering amplitude via 
\beq
f_\omega = [2 \sin^2 (\phi /2)]^{-n}  \kappa \sum_{m=-\infty}^\infty  X^{[n]}_m e^{i m \phi},
\eeq  
where
\beq
X^{[k+1]}_m = X^{[k]}_{m} - \tfrac{1}{2} \left( X^{[k]}_{m+1} + X^{[k]}_{m-1}\right), 
\eeq
and $X_m^{[0]} = e^{2 i \delta_m} - 1$. We find that $n=2$ % (\red{check}) 
is sufficient for an accurate numerical computation of $f_\omega$.

\subsection{Phase shifts}
Figure~\ref{phase_shifts} compares the phase shifts obtained via the numerical method, with 
 the analytic approximation, Eq.~(\ref{phaseshift-wkb}). The agreement is found to be good in the large-$|m|$ regime, as expected. The effect of absorption is clear in the low-$|m|$ regime, where $\left| \exp(2i\delta_m) \right| \ll 1$. In the intermediate regime, $m \sim \omega l_c^\pm$, the phase shift varies rapidly, due to large-angle scattering near the unstable orbits. 

\begin{figure*}[!tbh]
\centering
\includegraphics[scale=1]{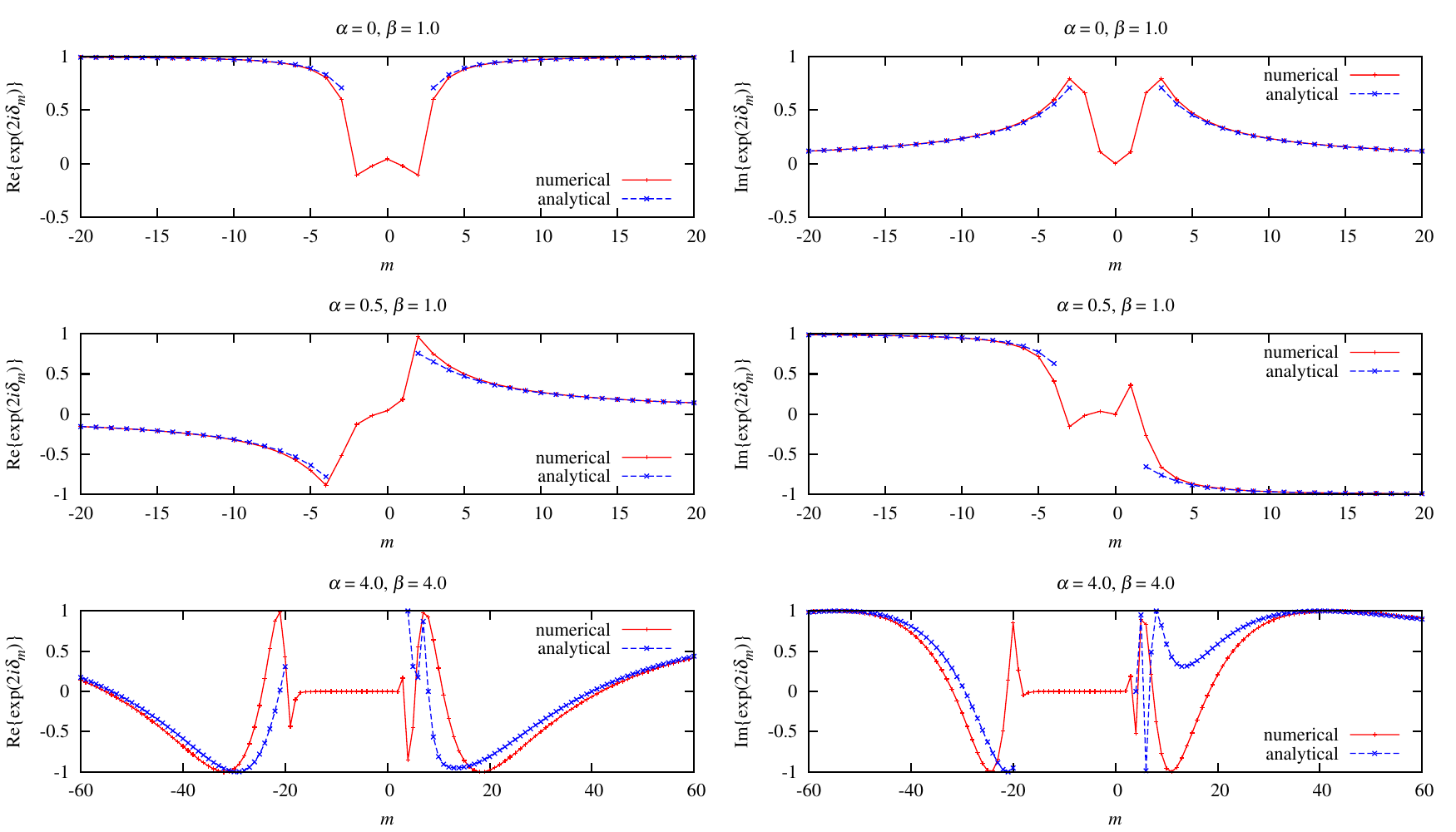}
\caption{Comparison of numerically-determined phase shifts [red, solid, crosses] with large-$m$ approximation [blue, dashed, x's]. The left-hand (right-hand) plots show the real (imaginary) part of $e^{2 i \delta_m}$. The numerical results for the phase shifts are in good agreement
with the analytical approximation, Eq.~(\ref{phaseshift-wkb}), for large
values of $|m| \gg \omega r_e$. In the regime $\omega l_c^{-} \lesssim m \lesssim \omega l_c^{+}$, where the approximation is not valid, the effect of absorption leads to the phase shift acquiring a significant imaginary part. %Note that the lowest plots use a different $x$-axis scale to the upper plots.
}
\label{phase_shifts}
\end{figure*}

With the numerical phase shifts, we may compute the
absorption and scattering length of the DBT vortex. 
The results for the absorption length can be found in Ref.~\cite{odc}; results for scattering length are presented in the next subsection. 

\subsection{Scattering length\label{subsec:scattering-length}}

\subsubsection{Low frequency scattering}

Figure~\ref{sca_comp} compares the low-frequency approximation for the scattering length, Eq.~(\ref{ana_sca}), with the 
numerical results, at 
$\beta = \omega r_h = 0.005$, $0.01$, and for $\alpha / \beta = C/D = 1.0$. The agreement is good in this regime, as expected. Moreover, the
agreement between analytical and numerical results improves as
the value of $\omega r_h$ decreases. This serves as a simple
consistency check on our analytical and numerical approaches.

\begin{figure}
%  \centering
 \includegraphics[width=8.6cm]{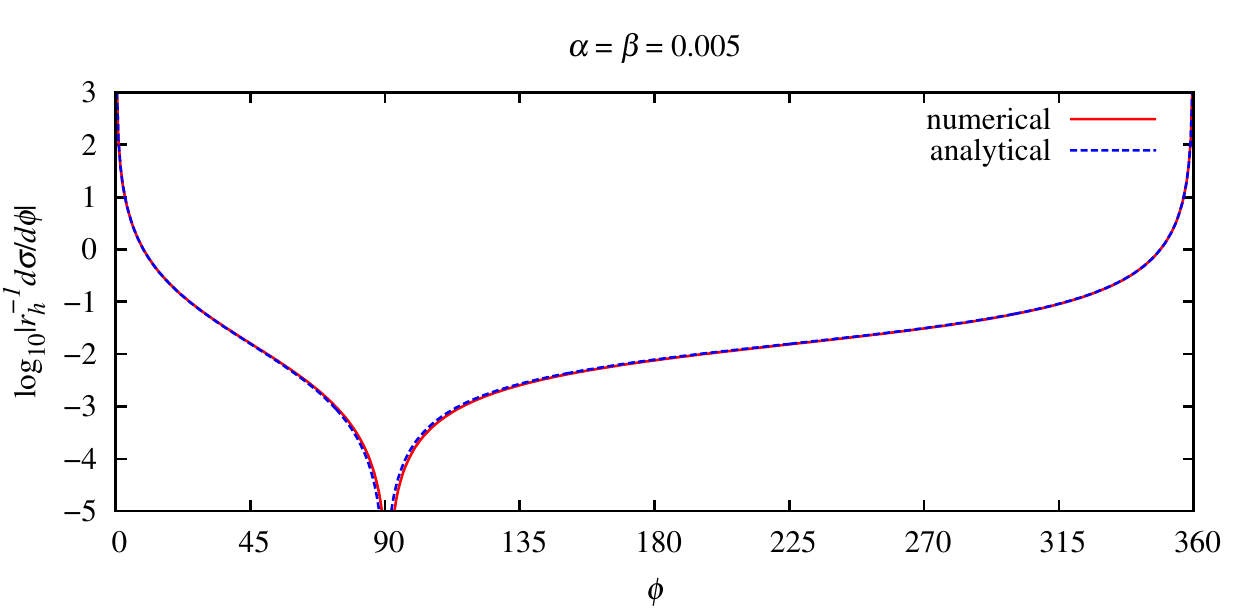}
 \includegraphics[width=8.6cm]{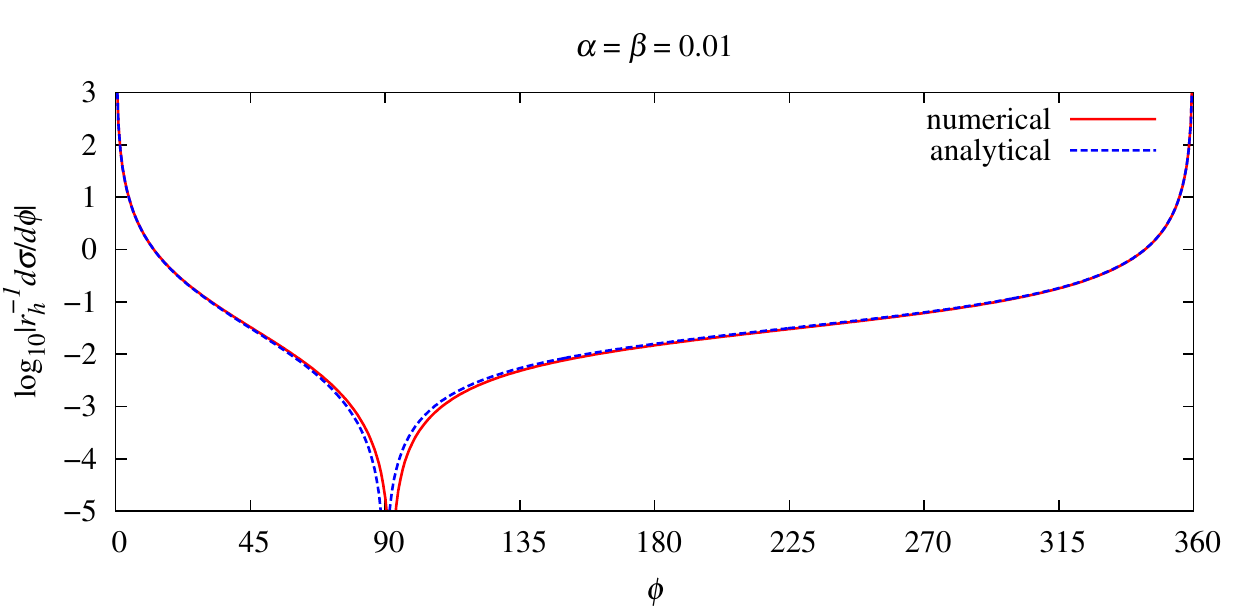}
 \caption{Comparison between the analytical (\ref{ana_sca}) and numerical results for
 the scattering length for $\alpha = \beta = 0.005, 0.01$.
 The results are in excellent agreement, particularly in the forward
 direction. The analytical model provides a better fit for smaller
 values of couplings $\alpha,\beta$.
}
 \label{sca_comp}
\end{figure}

\subsubsection{High frequency scattering}

Results for the scattering length for a selection of values of $C/D$
and $\omega r_h$ are presented in FIG.~\ref{fig:sca}. In all cases, the scattering length diverges in the limit $\phi \rightarrow 0$. In the regime $\Theta \ll \omega r_e$, $\Theta \ll 1$ (where here $\Theta = |\phi|$, $|2 \pi - \phi|$ as appropriate and $r_e$ is given by Eq.~(\ref{rh-re})), a simple approximation may be found by combining (\ref{weakfield}) and (\ref{classical-csec}) (or, equivalently, by applying the SPA (\ref{SPA}) to (\ref{amp}) with phase shifts (\ref{phaseshift-wkb})),
\beq
\frac{d \sigma}{d \phi} \sim \frac{\sqrt{3 \pi}}{4} \frac{r_e}{\Theta^{3/2}} .
\eeq
Asymmetric scattering effects (for $C \neq 0$) enter at the next order in the small-angle expansion.

The orbiting effect~\cite{Anninos} arises as regular interference fringes in the scattering length. %, induced by interference between contra-rotating rays. 
Figure \ref{fig:sca} shows that: (i) the angular width of the fringes is inversely proportional to the coupling $\omega r_e = \sqrt{\alpha^2 + \beta^2}$, as expected from Eq.~(\ref{Dphi}); (ii) as the circulation rate increases, the region of maximum interference is shifted away from the backward direction, in the same sense as the rotation of the DBT vortex.

\begin{figure*}
 \centering
 \includegraphics{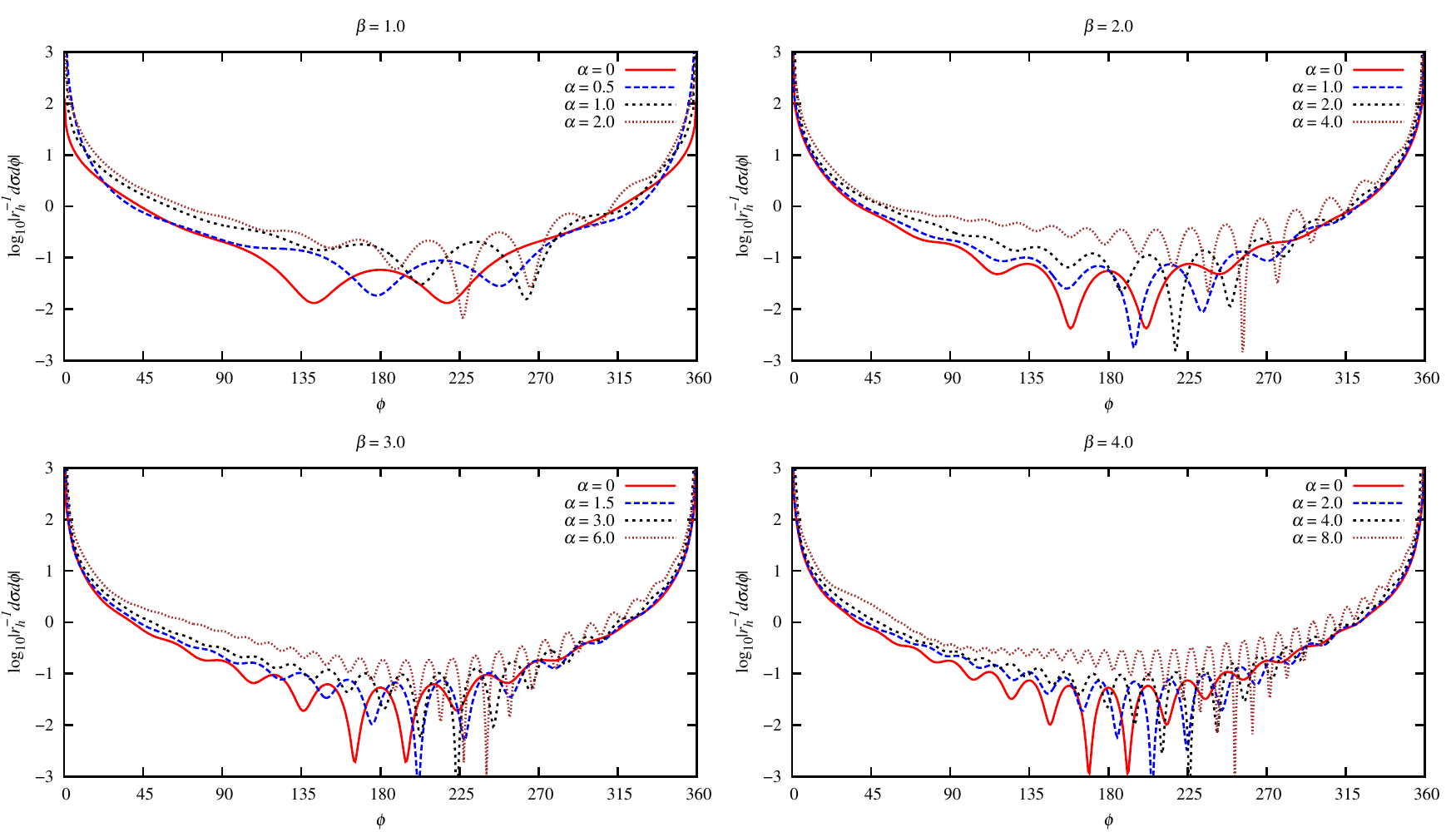}
 \caption{Scattering length of the DBT vortex for $\beta = 1.0, 2.0, 3.0, 4.0$, and $C/D=0, 0.5, 1.0, 2.0$. Note log-scale on the y-axis. Orbiting oscillations are present, with an (approximate) angular wavelength given by Eq.~(\ref{orbiting-wavelength}). The region of maximal interference is displaced in the co-rotating sense, as the value of $C/D$ increases.}
 \label{fig:sca}
\end{figure*}

The scattering length has some features in common with the
scattering cross sections of astrophysical black
holes~\cite{Futterman-1988}: for example, the interference fringes become narrower
as the coupling increases, and the scattering
length diverges as $\phi \to 0$. % This agrees with the fact that the effective potential in the weak-field limit falls as $1/r^2$~\cite{doc2}, which gives an infinite contribution to the forward direction scattering.
There is a key difference with respect to black-hole scattering, however, which is due to the nature of scattering in two (rather than three) spatial dimensions. Unlike for the Schwarzschild hole \cite{Futterman-1988} (and, for example, the canonical acoustic hole \cite{doc}), we do not observe a `glory' in DBT scattering.
A glory is a bright spot (or ring, for higher-spin fields) whose intensity increases as the wavelength decreases. A glory arises in three-dimensional scattering when geodesics from a one-parameter family (e.g.~a ring on the initial wave front obtained by rotating around the axis of symmetry in the Schwarzschild case) are focused onto a single point in the backward direction~\cite{Matzner-1985}. In the two-dimensional scenario this is not possible and hence the effect is absent. The magnitude of the large-angle scattering length remains small, even at high frequencies, and the orbiting oscillations may be a challenge to detect experimentally.

\subsubsection{Orbiting in the semi-classical approximation}

As shown in Sec.~\ref{sec:SCorb}, the orbiting oscillations may be related to geodesic scattering using the stationary-phase approximation. In essence, the peaks (troughs) arise due to the constructive (destructive) interference between contributions from the principal pair of contraorbiting geodesics. The scattering length is given by Eq.~(\ref{sig-semiclassical}), after inserting (\ref{sig-pm}), (\ref{eqI}) and (\ref{Dphi}). 

Figure \ref{fig-semiclassical} shows a comparison between the SC approximation and results from numerical summation of the partial-wave series, at $\beta = \omega r_h = 4$. The agreement is remarkably good across the full range of scattering angles, and for a wide range of circulation ratios $C/D$. We anticipate that the SC approximation could be further improved by including the (sub-dominant) contributions from rays which pass multiple times around the vortex. 

\begin{figure*}
\includegraphics[width=8.6cm]{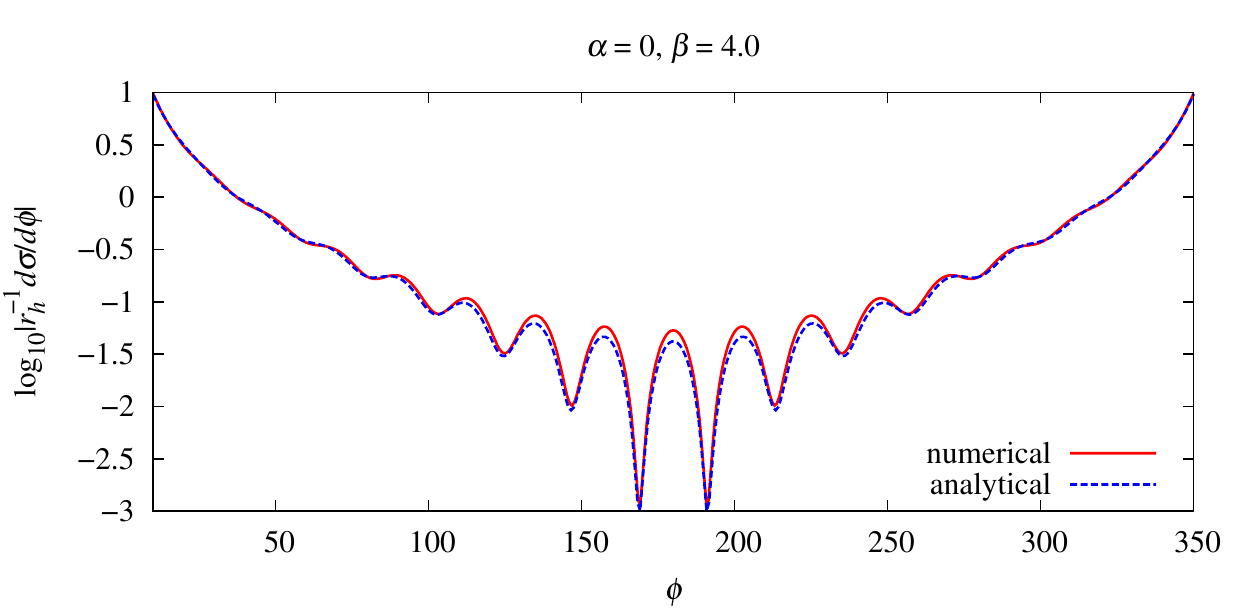}
\includegraphics[width=8.6cm]{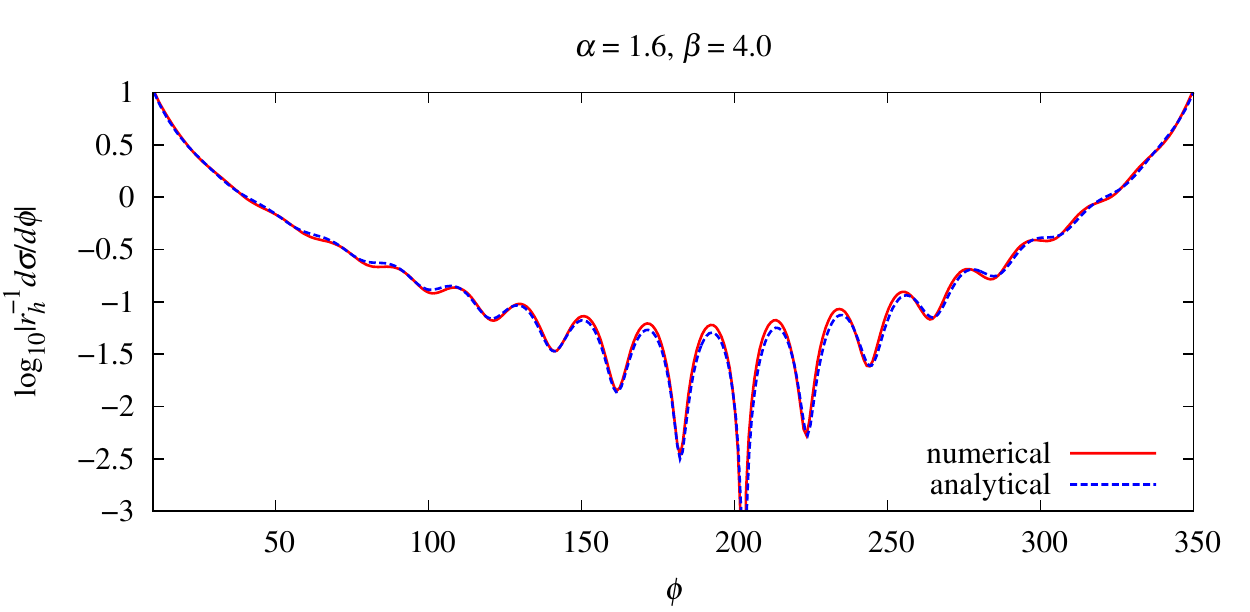}
\includegraphics[width=8.6cm]{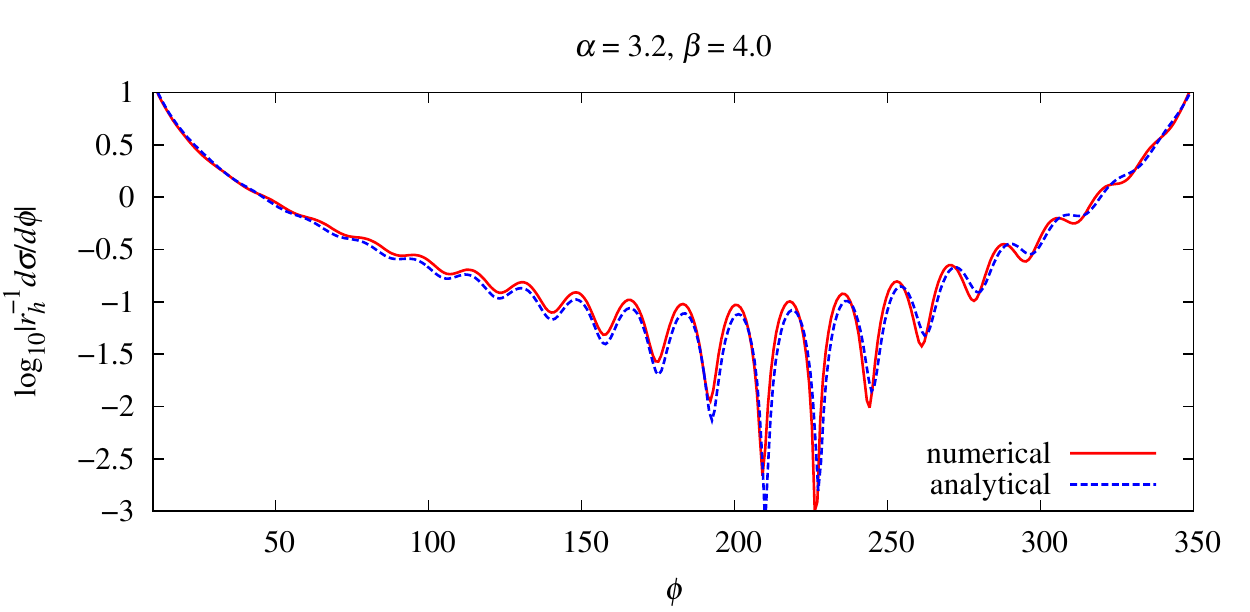}
\caption{Orbiting and the SC approximation. These semi-log plots show the partial-wave scattering length $r_h^{-1} d\sigma / d\phi$ as a function of scattering angle $\phi$, for frequency $\beta = \omega r_h = 4$ and circulation rates $C / r_h = 0, 0.4$ and $0.8$. The `numerical' result [red, solid line], found via numerical summation of series (\ref{amp}), is compared with the SC approximation [blue, dashed line], given by Eq.~(\ref{sig-semiclassical}). }
\label{fig-semiclassical}
\end{figure*}

\subsection{Wave scattering interference patterns\label{subsec:near-field}}

In quantum-mechanical scenarios, the wavefunction is inaccessible, and the experimentalist is restricted to sampling the probability density (i.e.~the square magnitude of the wave function) far from the scattering centre. In black hole scattering scenarios, the observer naturally resides far from the black hole, and thus the scattering amplitude $f_{\omega}$ and cross section $d \sigma / d \Omega$ remain the key quantities of interest. By contrast, in wave-scattering experiments conducted in the laboratory, one is able to observe the full interference pattern in the near-field; whereas extracting (e.g.) $d \sigma / d \phi$ in the far-field may be more difficult. 

To interpret the features of scattering in the near-field, let us draw comparisons with two canonical scenarios in 2D for which closed-form solutions are available, namely~(a) wave scattering by a hard circle and (b) scattering in the AB experiment. In scenario (a), 
\beq
\psi_{\text{circ.}} = \frac{1}{2} \sum_{m} i^m \left[ H^{(2)}_{m}(\omega r) - \frac{H^{(2)}_m(\tilde{\beta})}{H^{(1)}_m(\tilde{\beta})}  H^{(1)}_{m}(\omega r) \right] e^{i m \phi} ,
\eeq
where $\tilde{\beta} \equiv \omega a$ and $a$ is the radius of the hard circle, so that $\psi_{\text{circ.}} (r = a) = 0$ \cite{Lapidus} (see Ref.~\cite{Sakurai} for the equivalent 3D case). 
%
%\red{SRD: It seems to me more natural that the boundary condition should be $\partial_r \psi = 0$, so that the radial component of the flow velocity is zero -- how do we justify this boundary condition?} 
%\blue{ESO: If the circle is impenetrable, then there is no particle inside it. Therefore, it seems plausible to use a Dirchlet condition. We have the same for a impenetrable sphere. See Ref~\cite{Sakurai}, pg. 406, for instance.} 
%
In scenario (b), 
\beq
\psi_{\text{AB}} = \sum_m (-i)^\nu J_\nu( \omega r) e^{i m \phi}, \quad \nu \equiv \left| m + \tilde{\alpha} \right|.
\eeq

In FIG.~\ref{fig:hard-circle} we compare the interference patterns produced by planar-wave scattering by a non-circulating ($\alpha = 0$) DBT with scattering by a hard circle. In the hard-circle case (FIG.~\ref{fig:hard-circle}, right), a variety of effects are visible: (i) arcs of interference emanating from the scattering center, increasing in number with $\tilde{\beta}$, and prominent in both forward (downstream) and backward (upstream) directions, which resemble `seams of wavefront displacement'; (ii) a dappled interference pattern in a zone at intermediate angles, produced by spherical aberration; and (iii) a `shadow zone' in the forward direction. The shadow zone becomes particular prominent at large couplings $\tilde{\beta}$. The DBT pattern (FIG.~\ref{fig:hard-circle}, left) shares some of these characteristics, but does not exhibit interference arcs in the backward directions, nor the shadow zone. As $\beta$ increases, the arcs in the downstream zone become more distinct, and more numerous.  An 
obvious 
interpretation of the lack of upstream features in the DBT patterns of FIG.~\ref{fig:hard-circle} is that absorption of flux by the DBT acts to damp any direct back-scattering effects. 

\begin{figure*}[htb!]
\includegraphics[width=8.5cm]{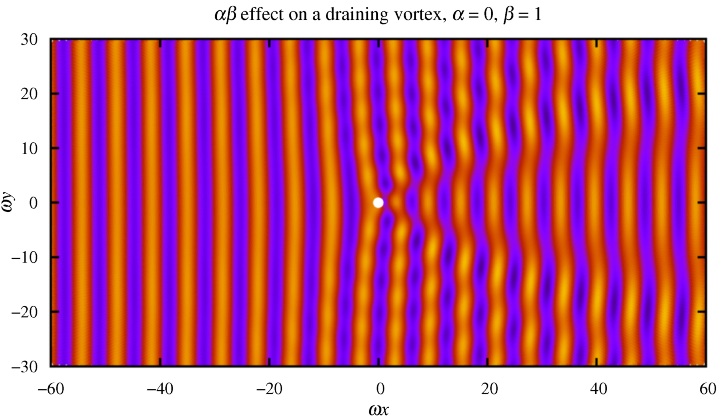}
\includegraphics[width=8.5cm]{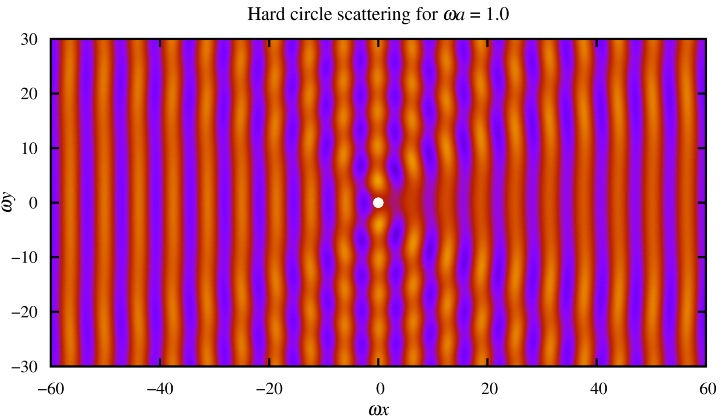}
\includegraphics[width=8.5cm]{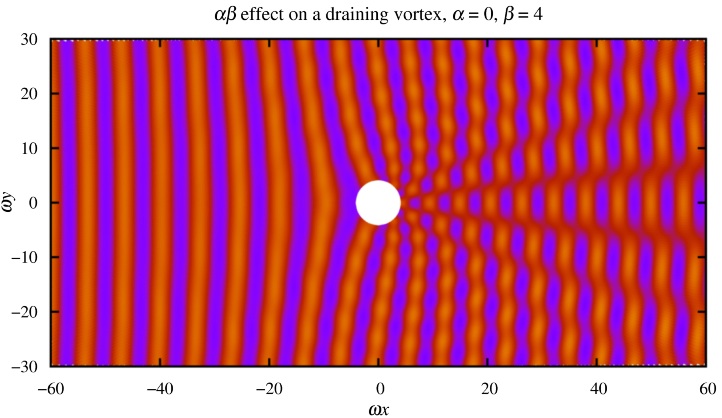}
\includegraphics[width=8.5cm]{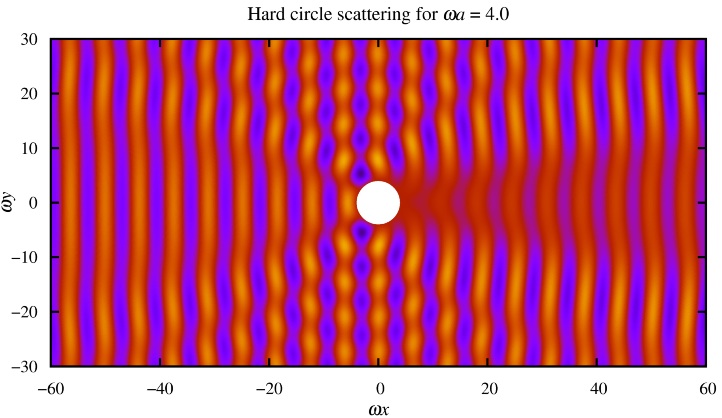}
\caption{Scattering in a non-circulating DBT (left) compared with scattering by a hard circle (right), at couplings $\beta = \tilde{\beta} = 1$ (upper) and $\beta = \tilde{\beta} = 4$ (lower). In each case, a planar wave is incident from the left.}
\label{fig:hard-circle}
\end{figure*}

In FIG.~\ref{ab_fig} we compare the scattering pattern from a circulating ($\alpha \neq 0$) DBT, with the AB scattering pattern. In the AB scattering, the scattering center splits the incident wavefront in two, and generates a path difference of $\tilde{\alpha} \lambda$ between the co- and counter-rotating segments. If $\tilde{\alpha}$ takes an integer value, then the wavefront segments reconnect smoothly on the opposite side of the vortex, as shown for $\tilde{\alpha} = 2,4$ in the middle and lower right plots of FIG.~\ref{ab_fig}. In addition, the scattering length in the AB effect is precisely zero when $\tilde{\alpha}$ is integer, according to Eq.~(\ref{AB_sca}). If $\tilde{\alpha}$ is non-integer then a `seam of reconnection' is generated in the forward direction, along which the phase is indeterminate. Figure~\ref{ab_fig} shows that, for $\alpha = 0.5$, this seam of reconnection is also present in the DBT scattering pattern (upper plots). However, for $\alpha = 2.0,4.0$, there is no smooth 
`reconnection' as in the AB case. By contrast, the DBT pattern exhibits seams in the forward direction, and spherical abberation, which give the pattern a very different character.

% Additional wave deflection and absorption occurs, and the scattering length is always bigger than zero.

\begin{figure*}[htb!]
\includegraphics[width=8.5cm]{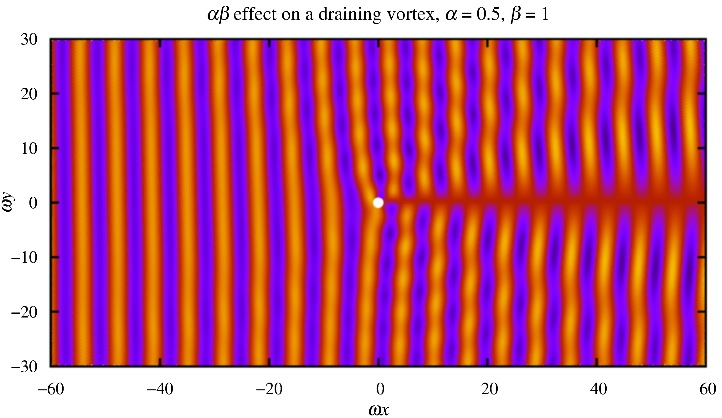}
\includegraphics[width=8.5cm]{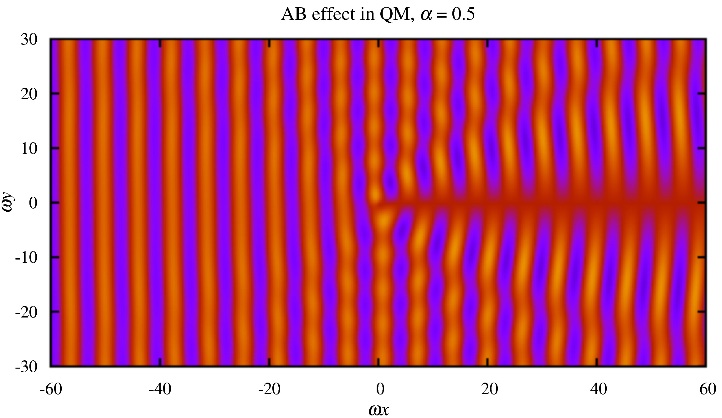}
\includegraphics[width=8.5cm]{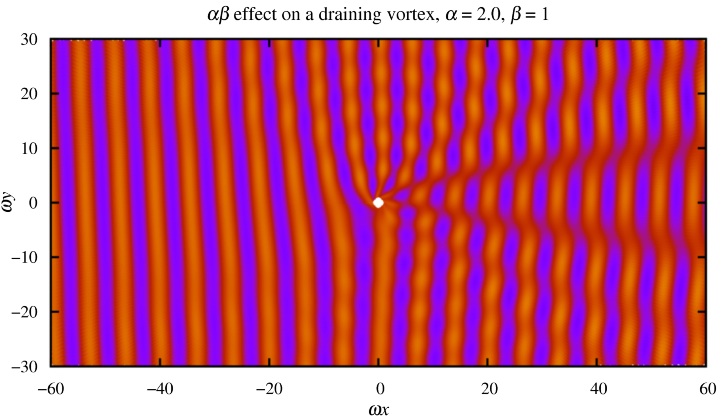}
\includegraphics[width=8.5cm]{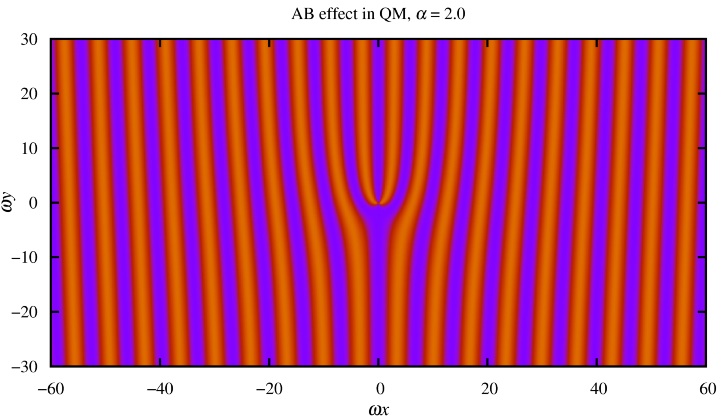}
\includegraphics[width=8.5cm]{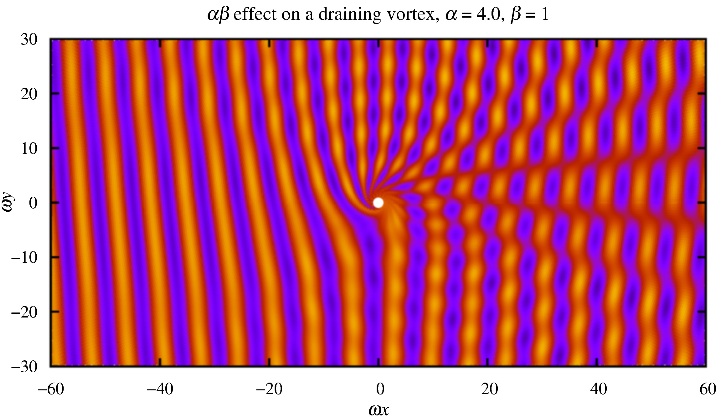}
\includegraphics[width=8.5cm]{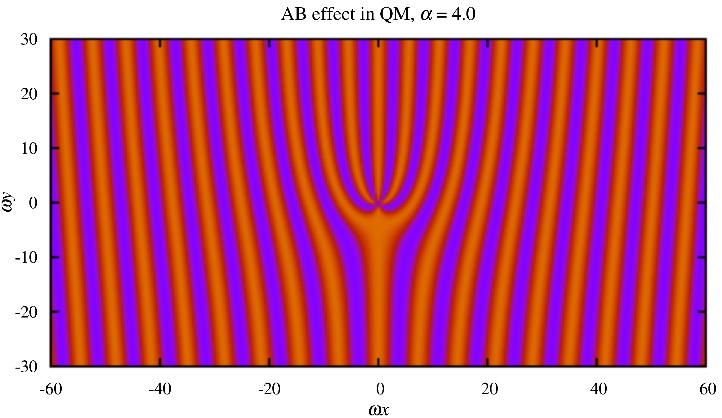}
\caption{Scattering in a DBT (left, the `$\alpha \beta$ effect' with $\beta = 1$) compared with scattering in the AB effect (right), at couplings $\alpha = \tilde{\alpha} = 0.5$ (upper), $2$ (middle) and $4$ (lower). }
\label{ab_fig}
\end{figure*}

\section{\label{sec:conc}Discussion and conclusion}

In this work we have examined, from a variety of perspectives, the scattering of a planar wave by a \textit{DBT vortex} -- a system which has been proposed as a fluid-mechanical analogue of a rotating black hole \cite{Unruh-Schutzhold-2002}. We have observed (i) `orbiting' oscillations in the scattering length, familiar from black-hole scattering studies \cite{Futterman-1988, Anninos}; (ii) at low frequencies, an AB effect modified by the absorption of flux in the circularly-symmetric ($m=0$) mode; and (iii) at higher frequencies, distinctive features of the near-field scattering pattern which arise from the interplay of rotation and absorption. As the scattering scenario is described by two couplings $\alpha$ and $\beta$ (see~Eq.~(\ref{alp-bet-def})), it seems natural to call this the $\alpha\beta$ effect \cite{doc2}.

The orbiting phenomenon (i), also known as `spiral scattering', also arises in black hole contexts; see Ref.~\cite{Anninos}. In essence, it is due to the interference of rays which pass in opposite senses around the scattering centre. In Sec.~\ref{subsec:high-freq} we showed that this simple interpretation leads on, via a semi-classical approximation, to a highly-effective approximation (see FIG.~\ref{fig-semiclassical}). In three-dimensional scenarios, orbiting is supplemented by an additional effect: the `glory'. A glory is a bright spot or ring in the forward- or backward-scattering directions, whose width (amplitude) decreases (increases) linearly with frequency \cite{Matzner-1985}. A glory is due to a one-parameter family of geodesics which scatter into a small solid angle. This is not possible geometrically in 2D; hence the effect is absent in the DBT case, and in surface-wave scattering more generally.

Our study has shown that an analysis of the scattering length (i.e.~the 2D version of the scattering cross section, defined as the intensity scattered into the far-field) illuminates only one aspect of the scattering scenario. Consider, as an example, the standard AB effect: the scattering length (\ref{AB_sca}) is precisely zero for integer values of $\tilde{\alpha}$; and yet the near-field interference pattern demonstrates topologically interesting features (see FIG.~\ref{ab_fig}), as shown experimentally by Berry \textit{et al.}~\cite{Berry}. Similarly, in the $\alpha\beta$ effect, the scattering length tells us relatively little about the near-field interference pattern, which will be directly observed in analogue experiments. For example, the orbiting effect seems to be too weak to be visible in the near-field (see FIGs. \ref{fig:hard-circle} and \ref{ab_fig}); instead the pattern is characterized by arcs in the forward direction. 

Studies of planar-wave scattering by black holes \cite{Matzner-1985, Futterman-1988, Andersson-Thylwe, Glampedakis-Andersson-2001, Anninos} usually take the view that the scattering cross section $d\sigma / d\Omega$ -- i.e.~the intensity per unit solid angle reaching a distant observer -- is the fundamental quantity of interest for any (future) experiment. We believe this perspective may be somewhat limiting, for several reasons. Firstly, in  gravitational-wave scattering, it is the wave \textit{amplitude} (rather than intensity) which determines its detectability. Secondly, constructive interference in the near-field may act to catalyze secondary effects. Thirdly, wave interference patterns in the vicinity of a black hole are surely of intrinsic interest, even if there is little scope for experimental investigations at present. We hope these arguments will motivate some further consideration of (e.g.) frame-dragging effects in the scattering by rotating black holes \cite{Glampedakis-Andersson-2001}.

Finally, there is the intriguing possibility that interference patterns like FIG.~\ref{fig:hard-circle} and FIG.~\ref{ab_fig} (left panels) could soon be observed in the laboratory. Here we can draw inspiration from the 1980s wavetank experiments of Berry \textit{et al.} \cite{Berry}, which showed how the AB effect is generated in a wavetank when planar waves impinge upon a vortex. Although Berry \textit{et al.}~did make use of a draining flow, their experiment was designed to investigate the AB effect, i.e. the effect of circulation (i.e. $\alpha$), rather than draining (i.e.~$\beta$) flow; nevertheless, additional features were visible (FIG.~4 in Ref.~\cite{Berry}). We believe there is now scope for a new analogue experiment to investigate planar-wave scattering on flows with circulating and draining components. Here the key challenge for an experimentalist seems to be to maintain the stability of the converging flow as it becomes supersonic, i.e.~in the vicinity of the ergoregion
(see Ref.~\cite{Andersen-etal}). Circumventing this 
problem may require creative approaches, or it may require the use of analogues in other media (e.g.~refractive materials, or Bose-Einstein condensates). Whichever route is taken, we hope that innovative experimental work will allow us to observe scattering patterns which resemble those of the  `$\alpha\beta$ effect', and which are similar in character to those that occur in the vicinity of astrophysical black holes.

\begin{acknowledgments}

The authors would like to thank Lu\'is C. B. Crispino for discussions and acknowledge Conselho Nacional de 
Desenvolvimento Cient\'\i fico e Tecnol\'ogico (CNPq) and 
Funda\c{c}\~ao de Amparo \`a Pesquisa do Estado do Par\'a (FAPESPA) 
for partial financial support. E. O. would like to acknowledge also partial financial support from 
Coordena\c{c}\~ao de Aperfei\c{c}oamento de Pessoal de N\'\i vel Superior (CAPES). 
S.~D.~thanks the Universidade Federal do Par\'a (UFPA) in Bel\'em for kind hospitality, 
and acknowledges financial support from the Engineering and 
Physical Sciences Research Council (EPSRC) under Grant No. EP/G049092/1.

\end{acknowledgments}

\appendix

\section{Semi-classical approximation: angular offset\label{appendix:offset}}
In this section, we obtain expressions for the angular offset $\chi$ which features in the SC approximation for orbiting oscillations, Eq.~(\ref{sig-semiclassical}) and (\ref{Dphi}). As our starting point, we use
\beq
\chi = - \Delta t / (4 r_e),
\eeq
where $\Delta t$ is the difference in the time taken for the (primary) co- and counter-orbiting rays that scatter precisely in the backward direction, i.e.~through an angle of $\pi$. To compute $\Delta t$ numerically (for a given $C, D$) we took two steps: (i) we found the specific angular momenta $l_\pi^{\pm}$ (for co- and counter-rotating orbits) which give scattering in the backward direction, $\Theta = \pi$, by solving Eq.~(\ref{scattering-angle}) numerically with the secant method; (ii) we computed the time difference numerically by calculating
\beq
\Delta t = 2 \lim_{\epsilon \rightarrow 0} \left[ \int_{\epsilon}^{u_0} \frac{\dot{t}}{\dot{\phi}} \left( \frac{d u}{d \phi} \right)^{-1} du   \right]^{+}_{-}
\eeq
where $\dot{t}$, $\dot{\phi}$, and $d u / d \phi$ are given in Eq.~(\ref{tdot-phidot}) and (\ref{orb-eq}), the root $u_0$ is determined from Eq.~(\ref{upoly}), and the square parantheses denote the difference between the integral evaluated on the co- and counter-rotating orbits. The limiting procedure is required because the integrals are formally divergent, whereas the difference is finite and well-defined. With this procedure, we obtain numerical data $\Delta t  / D \approx [0, -3.621, -7.235, -10.838, -14.428, -18.009 ] $ for $C/D = 0, 0.2, \ldots, 0.8, 1.0$. The relationship is almost linear-in-$C$, and well-fitted by $\Delta t \sim -18.1 C$. 

To understand the linear relationship, let us now consider a simple approximation in which the time difference is computed along the `critical' orbits (rather than along the neighboring scattered orbits). The polar angle $\phi$ along the orbit is given in terms of the radius $r = v^{-1/2}$ by
\beq
\phi =  \int_0^v \frac{ C+l - l(C^2 + D^2) v }{ 2 (1 - D^2 v)(1 - v / v_0) \sqrt{v} } dv,   \label{phic}
\eeq
where $1/v_0 = (r^\pm_c)^2 = |l_c^\pm| \sqrt{C^2+D^2}$, and $l$ takes the values $l_c^{\pm}$ [given in Eq.~(\ref{crit-orb-params})]. The time difference between co- and counter-rotating orbits is given by
\begin{eqnarray}
\Delta t / 2 &=& \lim_{\epsilon \rightarrow 0} \left[ \int_\epsilon^v \frac{1/v - C l}{2 (1-D^2 v)(1 - v/v_0) \sqrt{v}} dv\right]^+_-   \label{dt1} \\
 &=& \left[ \int_0^v  \frac{ (v_0^{-1} + D^2 - C l ) - D^2 v / v_0 }{ 2 (1-D^2 v)(1 - v/v_0) \sqrt{v} }  \right]^+_-  + r_c^- - r_c^+ . \nonumber
\end{eqnarray}
After noting that $\phi = \pi$ for the back-scattered geodesic, we may then add a multiple of (\ref{phic}) to (\ref{dt1}), to simplify the form of the integral, leading to
\begin{eqnarray}
\Delta t / 2 &=& \pi (l_c^+ + l_c^-) - (r_c^+ - r_c^-)  \nonumber \\  && + D \left[ \left(1 - z^2 \right) \text{arctanh} \left( \frac{1}{z} \right) \right]^+_- ,
\end{eqnarray}
where $z = r_c^{\pm} / D$. It is then straightforward to show that, at leading order in $C$, our approximation for the time difference is
\beq
\Delta t = - 8C \left( \pi - \text{arctanh}(1 / \sqrt{2}) \right) \approx -18.08 C .
\eeq
This linear-in-$C$ relationship is powerful as it fits the numerical data well over a large region of parameter space.

\bibliographystyle{apsrev}

%\bibliography{year1}

\end{document}